\documentclass[twoside]{article}
\usepackage{amsmath,amsfonts,bm,amssymb}
\usepackage{wrapfig}
\usepackage{cite}
\usepackage{cmpj}

\issue{2007}{10}{2(50)}{129}

\title[Analysis of a scientific periodical: a case study]
{Towards journalometrical analysis of a scientific periodical: a
case study}
\author{O.Mryglod\refaddr{a1,a2}, Yu.Holovatch\refaddr{a2,a3}}
\addresses{\addr{a1}Lviv Polytechnic National University, 12 Bandery Str.,
79013 Lviv, Ukraine%
\addr{a2}Institute for Condensed Matter Physics of the National
Academy of
Sciences of Ukraine, \\1 Svientsitskii Str., 79011 Lviv, Ukraine%
\addr{a3}Institut f\"ur Theoretische Physik, Johannes Kepler Universit\"at Linz, 69
Altenbergerstr., 4040 Linz, Austria%
}
\date{Received May 29, 2007}

\begin{document}

\maketitle
\begin{abstract}
In this paper we use several approaches to analyse a scientific
journal as a complex system and to make a possibly more complete
description of its current state and evolution. Methods of complex
networks theory, statistics, and queueing theory are used in this
study. As a subject of the analysis we have chosen the journal
``Condensed Matter Physics'' (http://www.icmp.lviv.ua). In
particular, based on the statistical data regarding the papers
published in this journal since its foundation in 1993 up to now
we have composed the co-authorship network and extracted its main
quantitative characteristics. Further, we analyse the priorities
of scientific trends reflected in the journal and its impact on
the publications in other editions (the citation ratings).
Moreover, to characterize an efficiency of the paper processing,
we study the time dynamics of editorial processing in terms of
queueing theory and human activity analysis.


\pacs 02.10.Ox, 02.50.-r, 07.05.Kf, 89.75.-k

\end{abstract}

\section{Introduction}
\begin{figure}[h]
\centerline{\includegraphics[width=15cm]{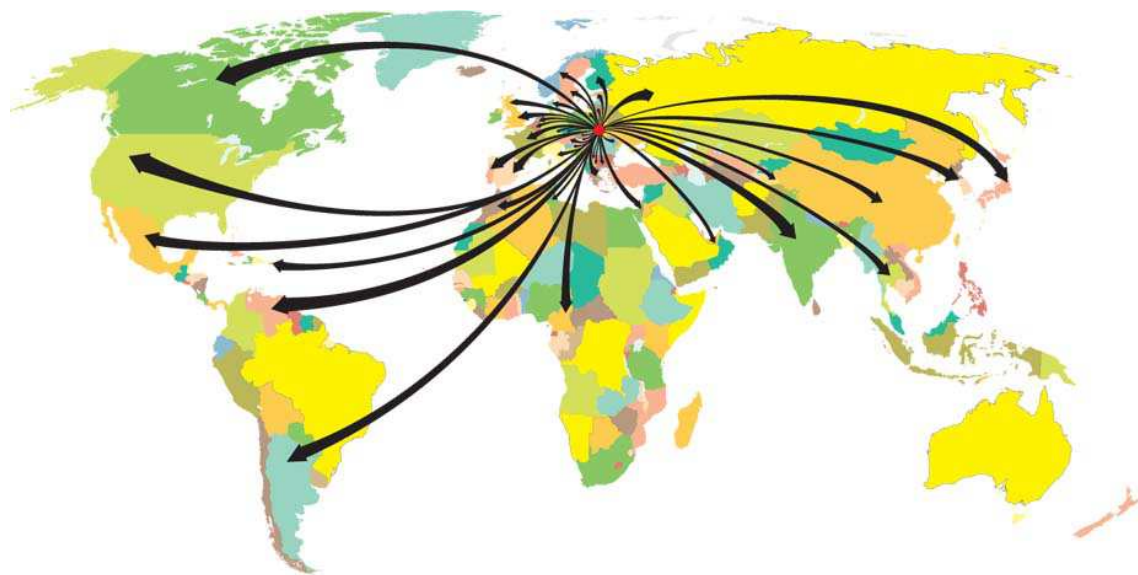}} %
\caption{An international collaboration of authors in ``Condensed
Matter Physics''. During the period of 1993--2007 the authors from
44 countries contributed to the journal.} \label{fig1_map}
\end{figure}

It is a honor and pleasure for us to contribute by our paper to
the jubilant fiftieth issue of the journal ``Condensed Matter
Physics'' (CMP) \cite{cmpref}. The history of this journal began
in 1993 when it was founded by the Institute for Condensed Matter
Physics of the National Academy of Sciences of Ukraine. Soon the
journal transformed into an international periodical, which is
currently recognized by the European Physical Society (since 2003)
and is covered by  ISI Master Journal List (since August, 2005).
Since the time of its foundation, 671 papers by authors from 44
countries have been published in the journal, see
figure~\ref{fig1_map}. The jubilee of the CMP is a good incentive
to present the results of the statistical analysis of its
publications, paying attention to their different features,
ranging from their content, collaboration trends of the authors to
the efficiency of the paper processing procedure. Recently, a new
term, {\em journalometry}, has appeared for the complex
quantitative analysis of  scientific periodicals \cite{Jeannin94}.
In particular, this complex approach makes it possible to take
into account different types of information about the journal,
ranging from the quantitative to the qualitative ones
\cite{Garfield90}.

Another reason for analysing the statistics of publications in a
scientific journal is that such an analysis permits to shed light
on different features of human activities as well as to quantify
them. To give an example, applying a complex network theory
\cite{networks} to the analysis of co-authorship of the papers
published, one deals with a collaboration network, a subject of
interest in social disciplines. Similar tools applied to the
analysis of the distribution of references that appear in the
papers published, lead to the so-called citation network, an
example of information networks. Moreover, as we shall discuss in
our paper, the analysis of distribution of the waiting times of
the papers submitted to the journal is useful in understanding the
origin of particular features of human dynamics
\cite{Barabasi05,Olivera05,Johansen04,Vazquez06,Stouffer06,MECO32}.
Although the analysed database allows us to make certain
conclusions about the statistical properties of the values
considered, one should be aware of the natural limitations imposed
by the finiteness of data set -- a typical situation, when a
particular periodical is studied.

The structure of our paper is as follows. In the rest of this
section we shall describe the database of CMP publications we
constructed. Section \ref{II} is devoted to the analysis of the
content of papers published in CMP. First, we address the authors
of the papers and construct the co-authorship network, measuring
its main characteristics. Then, we briefly discuss the main
thematic trends of the papers published and the way these papers
are cited in other periodicals. In section \ref{III} we approach
the analysis of the journal from another viewpoint. Here, the
subject of our analysis is not the content of a given paper but
rather the way the paper is processed by the editorial board. In
particular, we analyse the statistics of time intervals between
submission of a paper and its acceptance and interpret the
obtained results in terms of queuing theory. Conclusions and
outcome are given in section \ref{IV}.

\begin{figure}[h]
\centerline{\includegraphics[width=11cm]{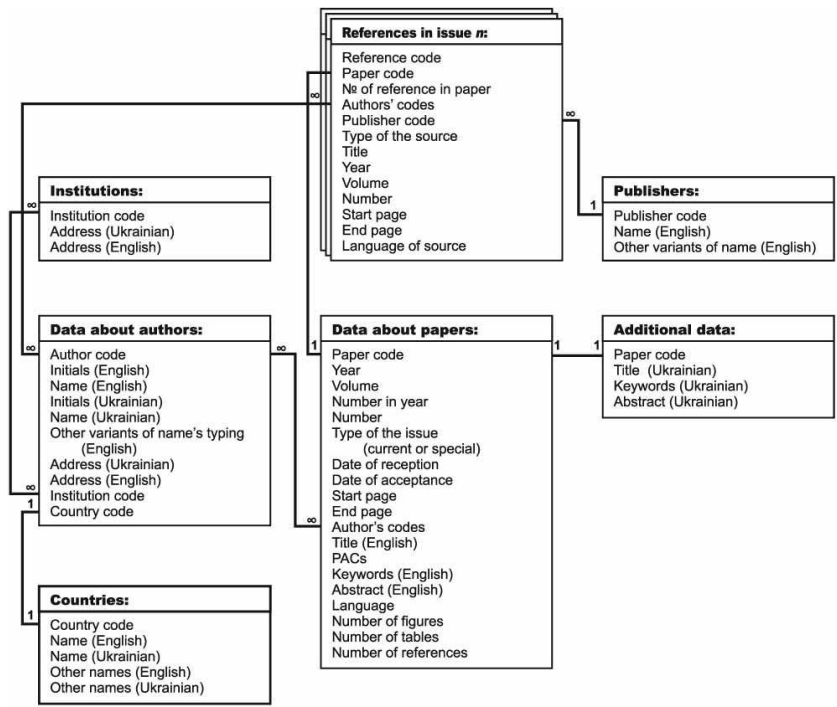}} %
\caption{The structure of ``Condensed Matter Physics'' journal's
database. } \label{fig2_DB}
\end{figure}

Before passing to a more detailed analysis of publications in CMP,
let us briefly describe the structure of the database we used as
well as display several general characteristics that follow from
the database analysis. The structure of the database is shown in
figure~\ref{fig2_DB}. It contains information about (i) the
authors and their affiliation, (ii) the content of the papers
published in CMP, as well as (iii) the data regarding the papers
cited in CMP. In what follows, we shall make use of the two first
parts of the database, the last one, (iii), will be considered
elsewhere.

\begin{figure}[!h]
\centerline{\includegraphics[width=7cm]{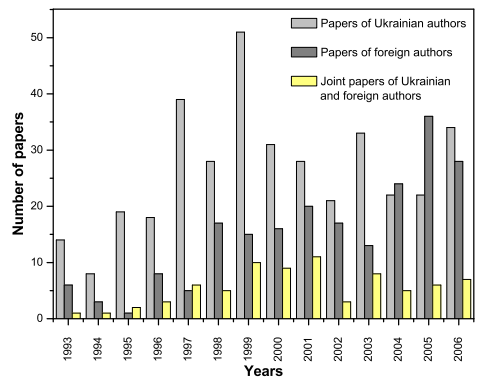}} %
\caption{The number of papers published in CMP by only Ukrainian,
only foreign author(s) and joint publications. }
\label{fig2_2_coop}
\end{figure}

During the period analysed, 892 authors published 671 papers in 49
issues of CMP, 477 authors being from Ukraine (where the
publishing institution is located) and 415 being from other
countries. The international cooperation increased with time. More
than a half of all papers during the last 3 years were written by
at least one foreign author (figure~\ref{fig2_2_coop}). The
decreasing rank of international co-authorship in CMP journal is
as follows: Germany (39 common papers), Poland (27), Japan (23),
Russia (22), USA (22), Austria (15), France(12) etc. However, the
data from the ISI database (Web of Science, \cite{ISI}) about the
external citation of CMP journal between 1993 and 2006
showed\footnote{The analysis was performed in October, 2006} that
the authors from other countries cited the papers from CMP in the
following decreasing order: USA (13.45\% of all citations),
Germany (12.48\%), France (9.75\%), Italy (8.58\%), Poland (8\%),
Austria (4.68\%), England (4.1\%), Russia (4.1\%) etc.
The maximum number of all coauthors in CMP per one person is 25
and the most active author has 29 publications here. 67 authors
had no coauthors in CMP at all.

Now, with the above described database at hand we shall perform a
more detailed analysis of the statistical parameters of the CMP
publications trying to find internal relationships between
different parameters and their time dynamics.

\section{Analysis of the content of the papers:
co-authorship and fields \\of research}\label{II}

In this section, we shall analyse several features connected with
the content of the papers published in CMP. Let us start from the
analysis of the authorship of the papers. To this end it is
reasonable to apply the tools of the complex networks theory
\cite{networks}, a field that originates from (and still may be
considered as a part of) graph theory. Complex networks were paid
much attention at the end of 1990-ies, when it appeared that many
network-like structures that exist in nature or result from human
activities possess remarkable properties which were not explained
within the then available mathematical framework. The most
striking ones were the so-called small-world and scale-free
features. In many networks the average distance between any pair
of nodes $\langle \ell \rangle$ appeared to be very small compared
to the network size $N$ (more precisely, the network is said to
possess small-world properties when $\langle \ell \rangle$ grows
with $N$ slower than $N^a,\, a>0$). Correspondingly, if the node
degree distribution $P(k)$ of a network is governed by a power
law, the network is said to be scale-free.

\begin{wrapfigure}{i}{0.38\textwidth}
\centerline{\includegraphics[width=45mm]{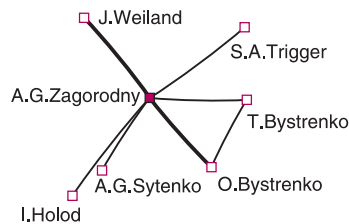}}%
 \caption{An example of a co-authorship graph. In this cluster a degree of
(number of links attached to) the central node is equal to six.
Lines of different thickness represent the number of common
papers.}
 \label{degree}
% \vspace{-3mm}
\end{wrapfigure}

The co-authorship network we are going to consider is one of the
examples of collaboration networks. In turn, the latter belong to
a special kind of social networks which represent human
collaboration patterns. Though social networks have a large
history, the collaboration networks have a more precise definition
of connectivity and much more data~\cite{Newman01_2}. A special
feature of collaboration networks is the possibility of
documenting all the facts of collaboration. In a collaboration
network, the particular individuals, groups of people or even
organizations can be represented by nodes, which are connected by
links (different types of interactions between people). An example
of a co-authorship graph from CMP is given in figure~\ref{degree}.
The nodes of the graph correspond to the authors and a link
between two nodes means that the two scientists represented have
coauthored a paper in CMP. Different ways of links weighting can
be performed: in non-weighed networks link means at least one
common paper for two authors; multiple lines or lines of different
thickness can represent the number of common papers, as it is done
in figure~\ref{degree}; in addition, the collaboration strength of
a link can be considered (for example, the strength of
collaboration between two authors is taken to be stronger than
between three authors)~\cite{Newman04}. The results of
investigations of scientific co-authorship networks have been
presented in numerous papers. Collaboration graphs for scientists
were constructed for a variety of fields based on the large
databases: MEDLINE (published papers on biomedical research), the
Los Alamos e-Print Archive (preprints in theoretical physics),
databases maintained by the Mathematical Reviews journal
(mathematical papers), NCSTRL (preprints in computer
science)~\cite{Newman01_2,Newman04}, SPIRES (papers and preprints
in high-energy physics)~\cite{Lehmann} etc. The sizes of these
databases range from 2~million papers to 13,000. The analysis
showed that co-authorship networks have the features of scale-free
small-worlds: they have a short mean distance between any two
nodes, a large clustering coefficient, and node-degree
distributions close to a power law. Actually, the distributions of
such networks are well fitted by  power laws with an exponential
cutoff. One of the possible explanations of this cutoff is the
finite set of data and the natural limitations on the active
working lifetime of a professional scientist~\cite{Newman01_2}.

\begin{figure}[!h]
\centerline{\includegraphics[width=12cm]{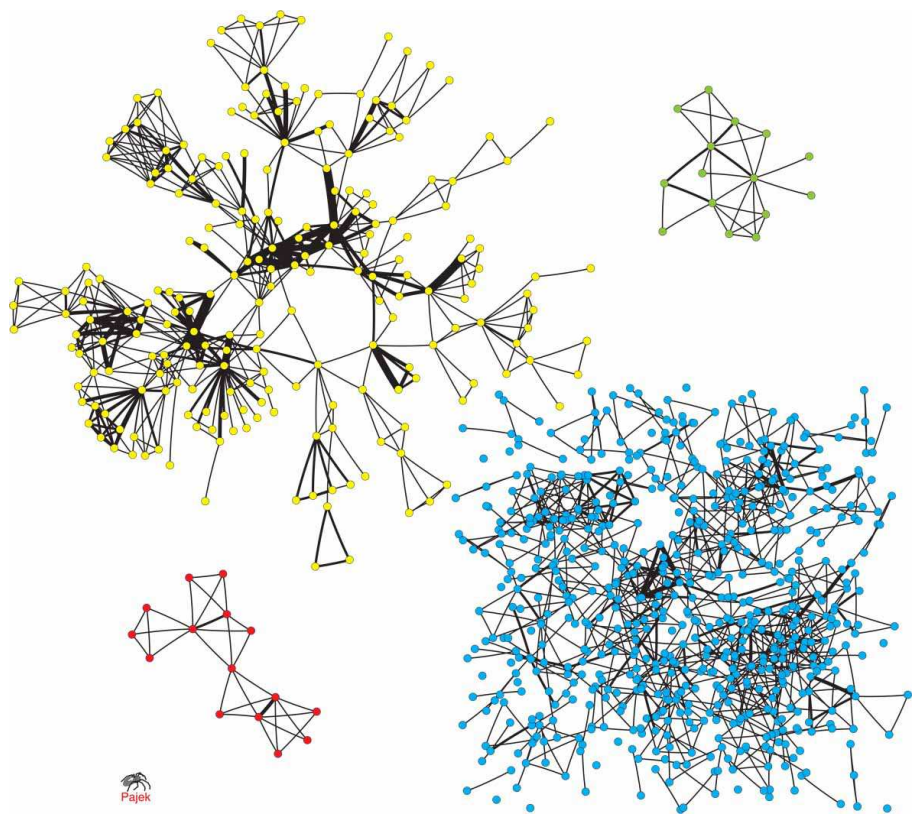}} %
\caption{Visualization of the co-authorship network (892 nodes) of
the CMP journal (1993--2007). Three different fragments can be
distinguished: the main cluster with 219 nodes (at the upper
left), the next-largest clusters with 15 nodes (top right and
bottom left). The rest of the nodes are collected on the right
below. The network was generated using Pajek network visualization
software~\cite{Pajek}.}%
 \label{fig3_netw}
\end{figure}

The general number of nodes in co-authorship networks varies
between scientific fields, their wide or narrow specializations
and periods of existence~\cite{Newman01_2}. On the other hand, the
average number of co-authors per one paper to a greater degree
depends on its type (experimental or theoretical). Naturally,
theoretical papers usually have a rather small numbers of
coauthors, compared with the experimental papers~\cite{Lehmann}.
Besides, in the experimental science the general number of
collaborators is larger. As regards the CMP, we can clearly see
the theoretical character of the journal: the average number of
co-authors per paper is equal to 2 and its maximum value is 8.

\begin{figure}[!h]
\centerline{\includegraphics[width=16.3cm]{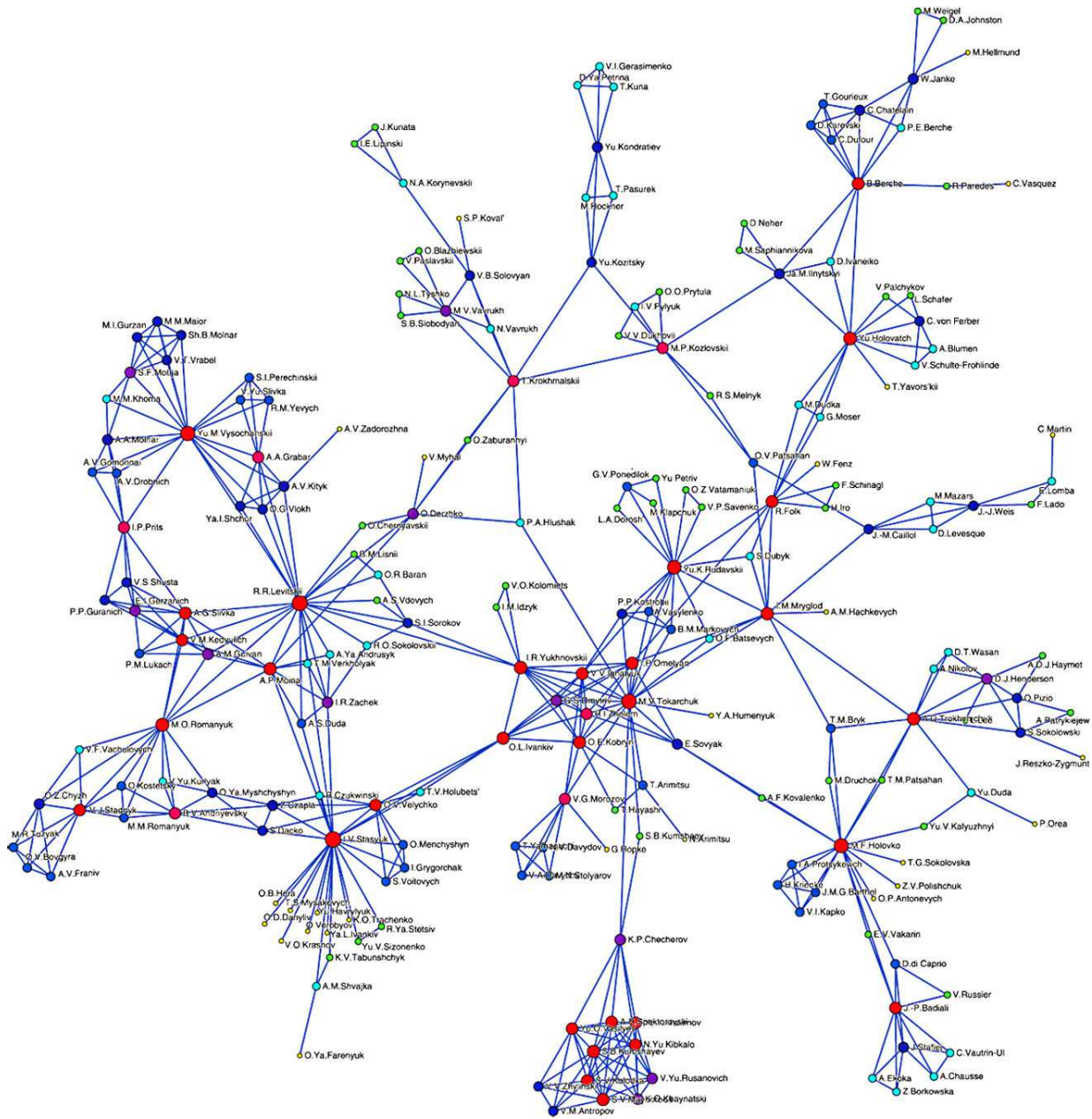}} %
\caption{The non-weighed main co-authorship cluster of the CMP
journal has 219 nodes (ge\-ne\-ral number of nodes is 892). Nodes
with different degree (different number of coauthors in CMP) are
denoted by circles of different colours and radii. The
visualization
was carried out using Himmeli software~\cite{Himmeli}.}%
 \label{main_cluster}
\end{figure}

The co-authorship network based on the relatively small database
consists of separate groups of connected nodes (clusters). These
fragments of the network can group the authors that work in the
same field of science. In each cluster every node can be reached
from any other node. With the growth of the size of the network,
separate clusters tend to join. At some moment, there appears a
giant cluster which includes almost all the nodes (a giant
component). The analysis of different co-authorship networks
showed that the main cluster includes approximately 80\% or 90\%
of all the nodes~\cite{Newman01_2,Newman04}. The existence of such
a cluster allows us to connect almost all the authors by one or
several chains of intermediate collaborators. The high level of
connectedness provides a fast spread of new scientific information
and shows intensive private interactions between scientists. It is
interesting that the second-largest connected cluster is far
smaller than the largest one. In figure~\ref{fig3_netw} we show
the co-authorship network of the CMP journal. The intensity of
each co-authorship, defined by the number of common papers, is
shown by lines of different thickness. The main co-authorship
cluster of the CMP journal is shown in figure~\ref{main_cluster}.

\begin{figure}[!h]
\centerline{\includegraphics[width=7cm]{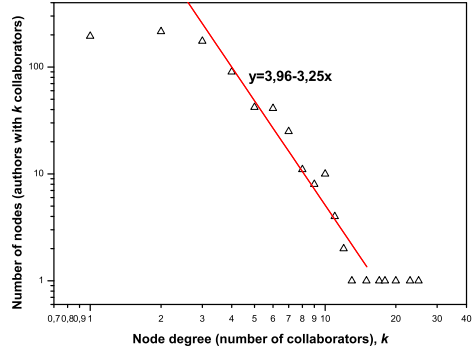}
}%
\caption{The node degree distribution of the co-authorship network
of the
CMP journal.}%
 \label{degreedist}
\end{figure}

Hight clustering is one of the main features of social networks.
The clustering of co-authorship networks greatly depends on the
number of papers with a few authors which automatically creates
cycles. Special parameters of a network may characterize its
clustering level. To proceed further, let us define the main
observables used for a quantitative description of
networks~\cite{networks}. As noted before, the degree $k_i$ of a
node $i$ is the number of the attached edges (see
figure~\ref{degree}). The mean node degree $\langle k\rangle$
characterizes the whole network:
\begin{equation}
\langle k\rangle=\frac{1}{N} \sum_{i=1}^{N}{k_i}\,,
\end{equation}
where the summation is performed over all $N$ nodes of a network.
For a non-weighed co-authorship network, the mean node degree is
the average number of coauthors of a particular scientist. The
node degree distribution $P(k)$ provides a probability for a node
to have a degree equal to $k$. The form of the node degree
distribution determines the network type. As noted above, the
network is said to be scale-free when its node degree distribution
follows a power-law:
\begin{equation}\label{distr}
P(k)\sim k^{-\gamma}, \qquad \gamma>0.
\end{equation}
In our case the node degree distribution is the distribution of
the number of collaborators of a scientist. The analysis of
co-authorship based on biomedical, physical and mathematical
databases showed that the corresponding distributions are
fat-tailed~\cite{Newman04}. The node degree distribution for the
CMP journal is shown in figure~\ref{degreedist}. In spite of a
small number of data points one definitely sees a tendency towards
a power law behaviour (\ref{distr}) with an exponent $\gamma\simeq
3.25$. Therefore, we conclude that the corresponding network is
scale-free.

The clustering coefficient $C_i$ of a node $i$ shows the
probability of the nearest neighbours of this node to be
connected. It is defined as:
\begin{equation}
C_i=\frac{2E_i}{k_i(k_i-1)}\,,
\end{equation}
where $E_i$ is the number of the existing connections between the
nearest neighbors of the node $i$ of $k_i$ degree. Respectively,
the mean clustering coefficient $\langle C \rangle$ characterizing
the whole network is defined as:
\begin{equation}
\langle C \rangle=\frac{1}{N}\sum_{i=1}^{N}{C_i}\,.
\end{equation}
The clustering reflects a special way of network correlation. The
clustering coefficient of a complete graph is equal to one,
whereas the clustering coefficient of a tree is zero. For the
Erd\"os-R\'enyi classical random graph (constituted by $N$ nodes
randomly connected by $L$ links) the clustering coefficient is
equal to:
\begin{equation}
C_{\mathrm{r}}=\frac{2L}{N^2}\,.
\end{equation}

Compared to a random graph, the scale-free networks have a
significantly larger value of the mean clustering coefficient
which proves their high correlation. In table~\ref{tab_results} we
give the mean value $\langle C \rangle$ of the CMP journal
co-authorship network compared to $C_{\mathrm{r}}$ of the random
graph of an equivalent size. The clustering coefficient, found for
the co-authorship network in physics (based on the publication in
Los Alamos E-print Archive \cite{losalamos}) between the years
1995 and 1999 is equal to 0.43~\cite{Newman04}; for its cond-mat
part (2000--2005) $\langle C \rangle \approx
0.73$~\cite{Cardillo06}. The mathematical co-authorship network of
the Mathematical Reviews journal has a clustering coefficient
equal to 0.15~\cite{Newman04}. The smallest value of $\langle C
\rangle$, 0.066, characterizes the biomedical field
(1995--1999)~\cite{Newman04}. The clustering coefficient of the
CMP journal is $0.607$ (see table~\ref{tab_results}).
\begin{table}[ht] \caption{The numerical characteristics of
the co-authorship network of the CMP journal. $N$: number of
nodes; $L$: number of links; $\langle k \rangle$,
$k_{\mathrm{max}}$: the mean and maximal node degree,
respectively; $\langle C \rangle$, $\langle C
\rangle/C_{\mathrm{r}}$: the mean clustering coefficient and the
ratio between clustering coefficients of a given network and of a
random graph of the same size; $\langle l \rangle$,
$l_{\mathrm{max}}$: the mean and maximal shortest path length.}%
\vspace{1ex}
\begin{center}
 \label{tab_results}
\begin{tabular}{|c|c|c|c|c|c|c|c|c|}
\hline Parameter&$N$&$L$&$k_{\mathrm{max}}$&$\langle
k\rangle$&$\langle C \rangle$&$\langle
C\rangle/C_{\mathrm{r}}$&$\langle
l\rangle$&$l_{\mathrm{max}}$\\\hline Value&
892&1300&25&2.915&0.607&185.3&4.783&10
\\\hline
\end{tabular}
\end{center}
\end{table}

\begin{figure}[!h]
\centerline{\includegraphics[width=7cm]{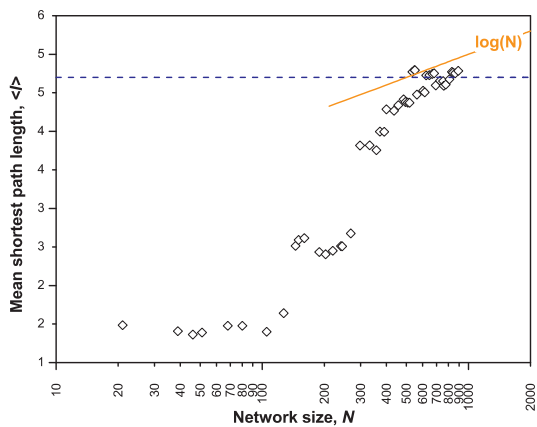}
}%
\caption{The change of the mean shortest path length $\langle l
\rangle$ with
an increase of the network size $N$ in the log-linear scale.}%
 \label{shortpath}
\end{figure}

In any connected fragment of the co-authorship network it is
possible to find the chains of intermediate collaborators between
any two authors. The results of calculations show a very small
lengths of the shortest paths between any two nodes: its average
value is close to 6~\cite{Newman01_2}. The length of the shortest
path $l_{ij}$ between the nodes $i$ and $j$ is equal to the
minimal number of links which should be passed to reach the $j$
from $i$. For a connected network, the mean  shortest path length
is defined as:
\begin{equation}
\langle  l\rangle=\frac{2}{N(N-1)}\sum_{i>j}{l_{ij}}\,.
\end{equation}
For well-connected networks the value of $\langle l \rangle$ is
not large: for the above mentioned database in physics
\cite{losalamos} $\langle l \rangle\approx 5.9$ for all data and
$\langle l \rangle\approx 6.4$ for its cond-mat
part~\cite{Newman01_2}. Measurements of the mean shortest path of
the papers submitted during 2000--2005 to the cond-mat part of the
Los Alamos E-print Archive resulted in the value $\langle l
\rangle\approx 3.62$ \cite{Cardillo06}. Our value of $\langle l
\rangle$ for CMP does not differ essentially (see
table~\ref{tab_results}). For small-world networks the value of
$\langle l\rangle$ scales logarithmically or slower with network
size~\cite{networks}. Figure~\ref{shortpath} shows how $\langle l
\rangle$ changes with an increasing size of the CMP co-authorship
network. Starting from a certain value of $N$ (approximately after
$N=550$) new nodes continue to appear but the mean distance
between them remains $\langle l \rangle\approx 4.7$.

Another value that characterizes the network size is the  maximal
shortest path length, $l_{\mathrm{max}}$. For our network, the
maximal shortest path $l_{\mathrm{max}}=10$ connects the nodes of
M.R.~Tuzyak and M.~Weigel, see table~\ref{tab_results}. Let us
note that the average value of $l_{\mathrm{max}}$ for
collaboration networks discussed above is between 20 and 30 and it
depends on the field of science~\cite{Newman04}; for physics
$\langle l_{\mathrm{max}} \rangle \approx 20$.

The problems connected with the shortest paths in collaboration
networks were the subject of the analysis in \cite{Newman04}. It
was shown that  on average 64\% of an individual's shortest paths
run through the best-connected of the nearest collaborators. This
fact shows the existence of very important nodes in co-authorship
networks, which can represent the most communicative and active
scientists. Another interesting parameter is ``transitivity''
which shows the probability that two coauthors of a scientist have
themselves coauthored a paper~\cite{Newman04}. In other words, if
two scientists have at least one common coauthor, they have a high
probability of becoming coauthors in future.

\begin{table}[!h]%
\caption{The top of the 10 most frequent PACS numbers in the papers of CMP journal.}%
\begin{center}
 \label{pacs}\footnotesize
\begin{tabular}{|l|c|c|}
 \hline \textbf{The name of the field}%
 $\vphantom{\frac{1^2}{2}}$%
 &\textbf{PACS number}&\textbf{Frequency}\\ \hline
\parbox{9cm}{Statistical physics, thermodynamics, and nonlinear
dynamical systems}&05&259
\\\hline  \parbox{9cm}{Structure of solids and liquids;
crystallography} &61 &154
\\\hline  Electronic structure of bulk
materials &71
 &139
\\\hline \parbox{9cm}{Equations of state, phase
equilibria, and phase transitions} &64
 &124
\\\hline \parbox{9cm}{Dielectrics, piezoelectrics, and
ferroelectrics and their properties} &77
 &116
\\\hline  Magnetic properties and
materials&75
 &73
\\\hline  Physical chemistry and
chemical physics&82
 &66
\\\hline
Superconductivity&74
 &58
\\\hline
\parbox{9cm}{Surfaces and interfaces; thin films and low-dimensional systems
(structure and nonelectronic properties)}&68
 &49
\\\hline
Physics of plasmas and electric discharges&52
 &39
\\\hline
\end{tabular}
\end{center}
\end{table}
A careful observation of the main co-authorship cluster of CMP
(figure~\ref{main_cluster}) makes it possible to single out its
strongly-connected fragments. The possible reason for the
existence of such connected groups is the common scientific
interest of these authors. In other words, one can visually
recognize the thematic trends of the CMP journal. Another
numerical data showing the priority fields of research in the CMP
is the statistics of PACS numbers. Table~\ref{pacs} represents the
top of the 10 most frequent PACS number in the papers of the CMP
journal. Finally, the data from the ISI database (Web of Science,
\cite{ISI}) regarding the external citations can show which
thematic fields of the CMP journal are most useful for the
scientists. The decreasing rank of subject categories that cited
CMP: physics, condensed matter (33.92\% of all citations),
physics, mathematical (14.43\%), physics, multidisciplinary
(14.04\%), physics, atomic, molecular \& chemical (12.87\%),
chemistry, physical (10.72\%), physics, applied (9.75\%), physics,
fluids \& plasmas (9.36\%), materials science, multidisciplinary
(7.21\%), electrochemistry (5.26\%), chemistry, analytical
(2.92\%), polymer science (2.53\%).

\section{Analysis of papers processing}\label{III}

\begin{figure}[!h]
\centerline{\includegraphics[width=9cm]{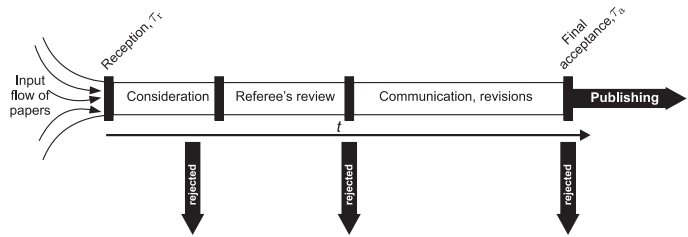}}
\caption{The schematic process of editorial processing of papers. $t$ shows the time arrow.}%
 \label{SchProcess}
\end{figure}

In this section, we approach the analysis of the CMP journal from
quite a different point of view. Here, the subject of the study
will be the way the papers submitted to CMP are processed by the
editorial board. A schematic process of editorial processing is
shown in figure~\ref{SchProcess}. Upon submission and
consideration by one of the editors, the paper is sent to the
reviewers, then revised (if necessary) and, finally, accepted. On
each of the above stages the paper may be rejected. However,
typically the information concerning the rejected papers is not
publicly available. Therefore, we define the waiting time of a
paper, $\tau_{\mathrm{w}}$, as the difference between the dates of
the paper final acceptance, $\tau_{\mathrm{a}}$, and the paper
reception $\tau_{\mathrm{r}}$:
\begin{equation} \label{tw}
\tau_{\mathrm{w}} = \tau_{\mathrm{a}} - \tau_{\mathrm{r}}\, .
\end{equation}
Both $\tau_{\mathrm{a}}$ and $\tau_{\mathrm{r}}$ are often
displayed in the paper. Therefore, the features to be analysed for
the CMP may be also checked for other periodicals \cite{MECO32}.
We shall be interested in the distribution $P(\tau_{\mathrm{w}})$
of the papers submitted to the CMP.

The distribution of waiting times during different kinds of human
activities (sending letters, e-mail communication, web-browsing,
loaning books in a library, etc.) has been a subject of recent
interest
\cite{Barabasi05,Olivera05,Johansen04,Vazquez06,Stouffer06}. In
particular, it was found that different forms of human activities
are characterized by a waiting time distribution in the form of a
power law:
\begin{equation}\label{wait}
P(\tau_{\mathrm{w}}) \sim \tau_{\mathrm{w}}^{-\alpha}\,.
\end{equation}%
Moreover, the dynamics of different processes appear to be
governed by different values of the exponent $\alpha$. The value
$\alpha=1$ governs the waiting time distribution for web-browsing,
e-mail communication and library loans
\cite{Johansen04,Vazquez06}; $\alpha=3/2$ for sending letters
(obtained concerning the analysis of correspondence of Einstein,
Darwin, and Freud) \cite{Olivera05,Vazquez06}; $\alpha=1.3$ for
stock broker activities~\cite{Vazquez06}. To explain this
phenomenon, a model of the queuing process based on the priority
principle has been used \cite{Barabasi05,Olivera05,Vazquez06}.
Note, however, the existing disagreement between the predictions
of \cite{Barabasi05,Olivera05,Johansen04,Vazquez06} on the one
hand, and the assumption of a log-normal distribution
$P(\tau_{\mathrm{w}})$ for e-mail communication patterns
\cite{Stouffer06}, on the other hand.

In a recent study \cite{MECO32} we proposed to analyse the forms
of waiting time distributions to characterize editorial processing
of scientific papers. To simplify the analysis we represented each
process of editorial consideration, referee review, communication
between the participants and the modification of materials being
one service action. The analysis of the waiting time distributions
of 2667 and 2692 papers published, respectively, in the journals
``Physica~A: Statistical Mechanics and its Applications'' (during
the period 1975 -- 2006) and ``Physica~B: Condensed Matter''
(during the period 1988 -- early 2007)
 lead to the conclusions about two possible forms of the
distribution ~\cite{MECO32}:
\begin{itemize}
\item power law with an exponential cutoff~\cite{Vazquez06}:
\begin{equation}
P(\tau_{\mathrm{w}}) \sim \tau_{\mathrm{w}}^{-b}\re
^{-\frac{\tau_{\mathrm{w}}}{\tau_0}}\,,\qquad
\tau_0>0,\quad b=1, \label{exp-pow}
 \end{equation}
where $\tau_0$ is the characteristic waiting time that depends on
the rates of submission and task execution \cite{Vazquez06};

\item log-normal~\cite{Stouffer06}:
\begin{equation}
P(\tau_{\mathrm{w}})\sim \frac{1}{\sqrt{2\pi}\omega
\tau_{\mathrm{w}}}\re ^{\frac{-\left[\ln
\left(\frac{\tau_{\mathrm{w}}}{\tau_{\mathrm{c}}}\right)\right]^2}{2\omega^2}}
\,, \qquad \omega>0 . \label{log-norm}
\end{equation}
where $\ln{\tau_{\mathrm{c}}}$ and $\omega$ are the mean and
standard deviation of the $\ln{\tau_{\mathrm{w}}}$.
\end{itemize}
Both distributions predict the same behaviour $\tau^{-1}$,
differing only in the functional form of the exponential
correction.

\begin{wrapfigure}{o}{0.48\textwidth}
\centerline{\includegraphics[width=6.3cm]{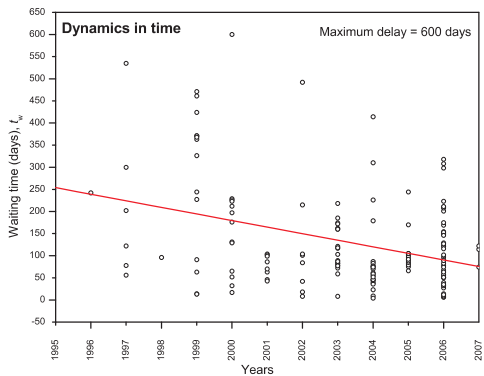}} %
\caption{Waiting time of the papers published in CMP between years
1995--2007. Different data points correspond to waiting times of different papers.
The mean waiting time (solid line) has a tendency to decrease.}%
 \label{Wait_dyn}
\end{wrapfigure}

\begin{figure}[!h]
\centerline{\includegraphics[width=7cm]{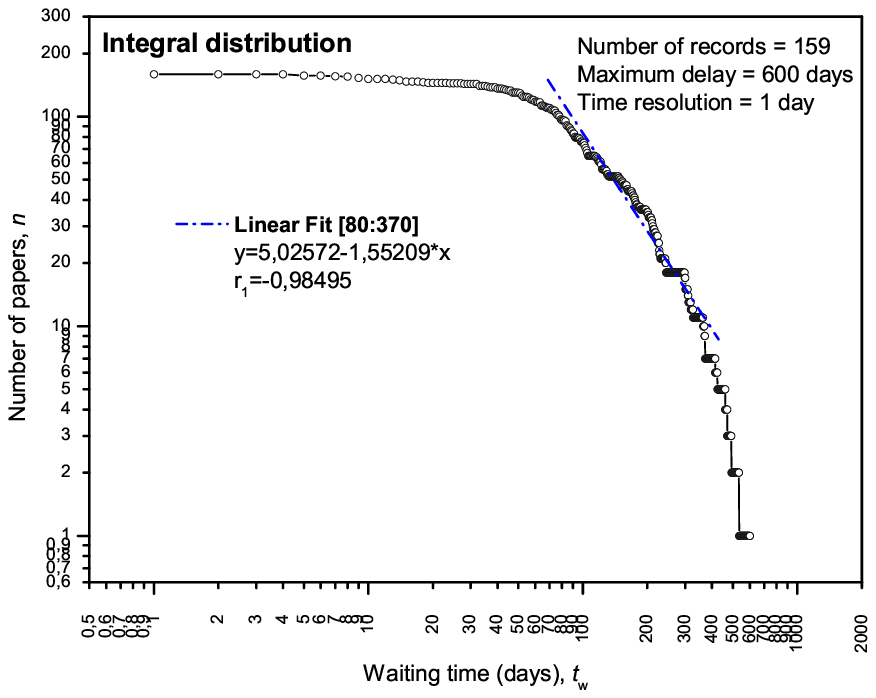}
\includegraphics[width=7cm]{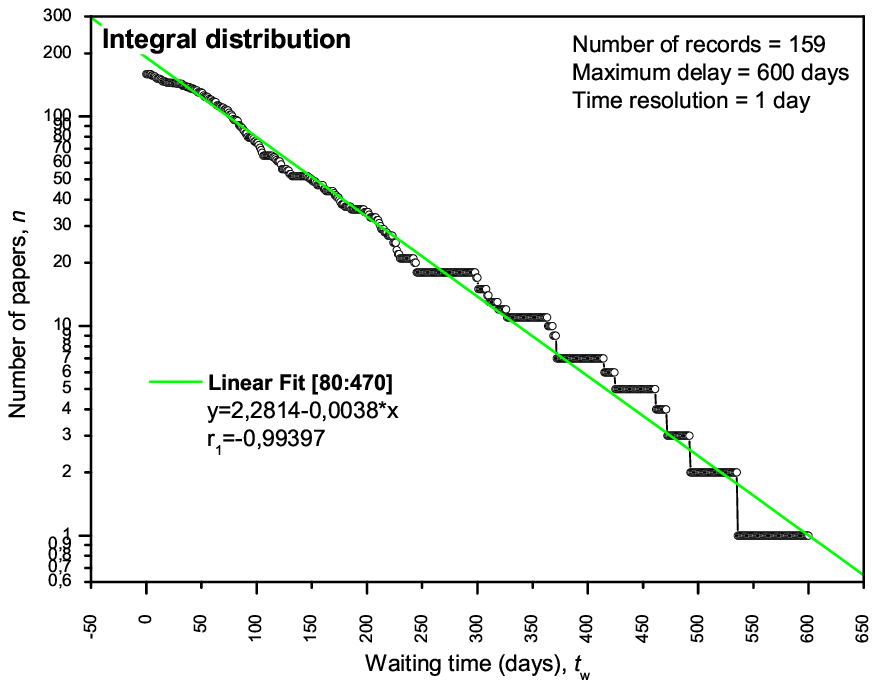}} %
\caption{Unnormalized integral distributions of waiting times: the
number of papers that have waited more than $\tau_{\mathrm{w}}$
days. Left plot: log-log scale, right plot: the same data in
log-linear scale.}%
 \label{Wait_IntLnLn}
\end{figure}

In figure~\ref{Wait_dyn} by the solid line we display the time
dynamics of the average waiting time of the papers submitted to
the CMP during the period  1995--2007. Different data points
correspond to waiting times of different papers. The maximal time
of the paper processing is equal to $\tau_{\mathrm{max}}=600$ days
(volume~9, No.~1(21), p.~175--182), the minimal one is equal to 4
days (volume~7, No.~4, p.~829--844; volume~7, No.~4, p.~845--858).
As one can see from the figure, the mean waiting time tends to
decrease. Unfortunately, only 159 out of all the papers published
in CMP contain the information on the dates of the papers final
acceptance, $\tau_{\mathrm{a}}$, and the paper reception
$\tau_{\mathrm{r}}$. Because of the poor statistics our results
show a significant data fluctuation. To obtain a smoother curve,
we analysed the corresponding integral distribution:
\begin{equation}\label{integr}
P^{\rm
int}(\tau_{\mathrm{w}})=\sum_{t=\tau_{\mathrm{w}}}^{\tau_{\mathrm{max}}}P(t).
\end{equation}
The corresponding curves (without normalization) are shown in
figure~\ref{Wait_IntLnLn} in the double logarithmic and log-linear
scales. A better linear fit (the absolute value of Pearson's
coefficient $r$ is closer to~1) on a log-linear scale suggests a
rather exponential behaviour of the integral distribution curve.

\newpage
\section*{Conclusions} \label{IV}
In this paper we have analysed the statistical properties of
different data that may be used in characterizing a scientific
periodical. In our ``case study'' we have chosen the journal
``Condensed Matter Physics''. Our analysis consisted of two parts.
In the first part we examined the features connected with the
content of the papers, whereas the subject of the second part was
the paper processing by the editorial board. By the content of the
paper we mean all the information contained in the text (authors,
affiliations, fields of research, etc). From different
characteristics that may be extracted in order to quantify the
content, we paid special attention to the authors and their
collaboration as well as to the main thematic trends of the
papers. To this end, we have analysed the CMP co-authorship
network using the methods of complex networks theory. Besides, we
made use of the ISI database to extract the data regarding the
citations of the papers published in CMP. In the second part,
where the paper processing was analysed, each paper was considered
without paying attention to its content. We were rather interested
in the statistics of waiting times of the submitted papers.

The main results obtained in our analysis are given in sections~2
and 3. To summarize some of them, let us mention that the
co-authorship network of the CMP journal and the main cluster of
authors were considered. We concluded about the scale-free nature
of this network as well as its great level of connectivity.
Moreover, a very positive tendency of the improving international
collaboration is observed. External ratings and evaluations should
be also taken into consideration in the full complex journal
analysis. This information can be obtained from different
international information services and databases such as: Thomson
ISI (Web of Science)~\cite{ISI}, Scopus~\cite{Scopus} and others.
Mainly, they include the external citation data which can be the
base of the most known quantitative criteria for the evaluation of
scientific efficiency. Another result is connected with the study
of the editorial board's activities. The editorial processing of
incoming papers can be considered as a process of human activity.
Analyzing the statistics of waiting times of the papers we can
find a promising decrease of this parameter for CMP journal during
the observed period.

Of course, our study does not cover the analysis of all the data
that may be considered on the basis of the database constructed.
To give an example, we did not touch upon the analysis of citation
and co-citation networks, etc. Nevertheless, we think that our
study might be useful as far as the external evaluation of the
journal is considered. Moreover, some of the numerical data given
in our paper can help to evolve a strategy to improve the work of
the editorial board.

\newpage
\section*{Acknowledgements}
O.M. thanks the Johannes Kepler Universit\"at  Linz (Austria) for
the possibility of getting information from the ISI database and
to Ihor Mryglod for the help rendered. We acknowledge useful
discussions with Christian von Ferber and Reinhard Folk.


\begin{thebibliography}{99}

\bibitem{cmpref} The ISSN number of the journal ``Condensed Matter Physics''
is ISSN 1607--324X and the http address reads:
http://www.icmp.lviv.ua/journal/index.html

\bibitem{Jeannin94}
Jeannin P., Devillard J. Towards a demographic approach to
scientific journals, Scientometrics, 1994, \textbf{30}, 83--95. Note
however that this reference analyses an evolution of a set of
scientific journals, whereas in our analysis we are rather
interested in 'measuring' a single journal.

\bibitem{Garfield90} Garfield~E. How ISI selects journals for coverage:
Quantitative and qualitative considerations. Current contents,
1990, No.~22, 5--13.

\bibitem{networks}
Albert~R., Barab\'asi~A.-L. Statistical mechanics of complex
networks, Rev. Mod. Phys., 2002, {\bf 74}, 47; Dorogovtsev~S.N.,
Mendes~J.F.F. Evolution of networks, Adv. Phys., 2002,  {\bf 51},
1079; Newman~M.E.J. The Structure and Function of Complex
Networks, SIAM Review, 2003, {\bf 45}, 167; Dorogovtsev~S.N.,
Mendes~S.N. Evolution of Networks: From Biological Nets to the
Internet and WWW. Oxford University Press, Oxford, 2003;
Holovatch~Yu., Olemskoi~O., von~Ferber~C., Holovatch~T.,
Mryglod~O., Olemskoi~I., Palchykov~V. Complex networks, J.~Phys.
Stud.,  2006, \textbf{10}, in press (in Ukrainian).


\bibitem{Barabasi05}Barab\'{a}si~A. The origin of bursts and heavy tails in human
dynamics, Nature, 2005, \textbf{435}, 207--211.

\bibitem{Olivera05}Olivera~J.G., Barab\'{a}si~A. Darwin and Einstein
correspondence patterns, Nature, 2005, \textbf{437}, 1251.

\bibitem{Johansen04} Johansen A. Probing human response times, Physica A, 2004, 338, No.~1--2, 286--291.

\bibitem{Vazquez06} Vazquez A., Oliveira J.G., Dezso Z., Goh K.-I., Kondor I.,
Barabasi A. Modeling bursts and heavy tails in human dynamics. Phys.
Rev.~E, 2006, \textbf{73}, 036127.

\bibitem{Stouffer06}Stouffer~D.B., Malmgren~R.D., Amaral~L.A.N.
Log-normal statistics in e-mail communication patterns,
arXiv:physics/0605027, v1 3 May 2006.

\bibitem{MECO32} Mryglod~O., Holovatch~Yu. The patterns of natural
human dynamics at tasks execution in queue: statistics of waiting
times for scientific articles. -- In: MECO32, Poland, L{\c{a}}dek
Zdr\'oj, April 16--18, 2007, and unpublished.

\bibitem{ISI} The product of Thomson Scientific, Web of Science:  http://scientific.thomson.com/products/wos/

\bibitem{Newman01_2}  Newman~M.E.J. The structure of scientific
collaboration networks, Proc. of the Nat. Acad. of Sciences, USA,
2001, \textbf{98}, No.~2, 404--409.

\bibitem{Newman04}Newman~M.E.J. Coauthorship networks and patterns of
scientific collaboration, Proc. of the Nat. Acad. of
Sciences, USA, 2004, \textbf{101}, 5200--5205.

\bibitem{Lehmann}Lehmann~S. Spires on the Building of Science: Complex
Networks and Scientific Excellence. Cand. Scient. Thesis, 2003.
The Niels Bohr Institute.

\bibitem{Pajek}A. Vlado. Pajek: Program for large network analysis.
http://vlado.fmf.unilj. si/pub/networks/pajek/.

\bibitem{Himmeli}V.-P. Makinen. Himmeli: Graph drawing software.
http://www.artemis.kll.helsinki.fi/himmeli/.

\bibitem{losalamos} ArXiv.org: open access to 420,559 e-prints in
Physics, Mathematics, Computer Science and Quantitative Biology,
http://arxiv.org/

\bibitem{Cardillo06}Cardillo~A., Scellato~S., Latora~V. A
topological analysis of scientific co-authorship networks,
Physica~A, 2006, \textbf{372}, 333--339.

\bibitem{Scopus}http://www.scopus.com/



\end{thebibliography}
\end{document}